\documentclass{INTERSPEECH2023}

\interspeechcameraready

\title{Voice Passing : a Non-Binary Voice Gender Prediction System  for evaluating Transgender voice transition}
\name{David Doukhan$^1$, Simon Devauchelle$^1$, Lucile Girard-Monneron$^2$, Mía Chávez Ruz$^3$, V. Chaddouk$^3$, Isabelle Wagner$^2$, Albert Rilliard$^{4,5}$}

\address{
  $^1$Institut National de l'Audiovisuel (INA), France; 
  $^2$Hôpital Tenon, AP-HP, France;
  $^3$Independent;
$^4$Université Paris Saclay, CNRS, LISN, France;
$^5$Universidade Federal do Rio de Janeiro, Brazil}

\email{\{ddoukhan, sdevauchelle\}@ina.fr, lucile.monneron@aphp.fr, mia.chavezruz@gmail.com, vancdk@pm.me, isabelle.wagner@aphp.fr, rilliard@lisn.fr}

\begin{document}

\maketitle
 
\begin{abstract}
This paper presents a software allowing to describe voices using a continuous Voice Femininity Percentage (VFP). This system is intended for transgender speakers during their voice transition and for voice therapists supporting them in this process. A corpus of 41 French cis- and transgender speakers was recorded. A perceptual evaluation allowed 57 participants to estimate the VFP for each voice. Binary gender classification models were trained on external gender-balanced data and used on overlapping windows to obtain average gender prediction estimates, which were calibrated to predict VFP and obtained higher accuracy than $F_0$ or vocal track length-based models. Training data speaking style and DNN architecture were shown to impact VFP estimation. Accuracy of the models was affected by speakers' age. This highlights the importance of style, age, and the conception of gender as binary or not, to build adequate statistical representations of cultural concepts.

\end{abstract}



\section{Introduction}


Care of transgender people offers, among other things, support for the modification of the gender perceived through their voice, a main component of identity - particularly of gender identity~\cite{gray2019transgender}.
During female-to-male transitions, the lowering of a voice's fundamental frequency ($F_0$) is relatively easy to obtain by taking testosterone, which produces a lengthening of vocal folds and a masculinization of voice. Conversely, raising one's vocal pitch so the voice is perceived as female is more complex, as taking female hormones does not influence the phonatory structure after puberty for males \cite{Fugain_2019,schmidt2018voice}.
However, voice's $F_0$ is not the only criterion for identifying voice gender. 
Voice quality (and typically vocal tract resonances) is also an important determinant, while prosody, speech rhythm, and vocabulary intervene as secondary cues~\cite{murry1980multidimensional,hillenbrand2009role}.
Beyond personal training, transgender persons are offered care by voice therapists and, if required, two types of vocal surgery that aim a raised voice pitch: cricothyropexy and Wendler glottoplasty \cite{VanBorsel_VanEynde_DeCuypere_Bonte_2008,Mastronikolis_Remacle_Biagini_Kiagiadaki_Lawson_2013} 
A recurrent question is how to evaluate this work and its outcome? (i.e., how do we estimate the gender that'll be perceived from a voice in a given language and culture?) Most available software (e.g., EvaF~\cite{evaf}
or VoiceUp~\cite{voiceup}) proposing the evaluation of voice masculinity or femininity for non-expert essentially use voice $F_0$.
While this measurement alone does not capture gender perception (an extra-high $F_0$ may correspond to a falsetto voice, or a very low $F_0$ to a partial laryngectomy...), let alone not being tuned to cultural variation  \cite{vanBezooijen_1995}. 

This paper describes the setup and evaluation of a tool trying to close the gap between the reality of voice therapy practice and gender perception. A program allowing transgender persons to train their voice and measure their progress, and allowing voice therapists to evaluate and develop their techniques with a tool adapted to their daily needs.
This tool thus (i) shall take into account the complex characteristics of a voice (not only $F_0$); (ii) shall return a proportion of masculinity/femininity so transgender persons may adapt the output to their own want.
To develop this service, machine learning (ML) algorithms were trained to evaluate the voices' gender, and their outputs were tuned to the perceptual evaluation of a corpus of individual voices by naive French listeners. This perceptual evaluation and the setup of these algorithms, with their performance evaluation, is the topic of this paper. This study focuses on gender perception within the French culture and language.


\section{Related Work}


Earlier gender prediction systems were based on LPC analysis~\cite{childers88}, MFCC gender-dependent HMM phone recognizer~\cite{lamel1995phone}, or Mel bands and pitch estimation HMM~\cite{543213}.
This task was defined as a binary classification problem associated with high accuracy estimates ($> 95\%$), often considered as solved.
However, the reported performances were not necessarily comparable since accuracy depends on e.g., corpora, sample duration, speech transcript, speaking style, speaker age, and language.
Recent studies, using pre-trained Transformer-based acoustic features~\cite{lebourdais2022overlaps} or Convolutional Neural Networks (CNN) trained on Mel bands~\cite{doukhan2018open,bensoussan2021deep}
, reported accuracy metrics above 90\% on fixed-length speech samples (2 seconds, 680 ms, 30 seconds), but also gender classification biases defined as  accuracy differences between female and male speakers. 
For transgender voices, classification systems used three gender categories: a male, female, and transgender system was fitted to recordings of cis- and transgender (male \& female) speakers, with an accuracy of 83\%~\cite{yasmin2022rough}.
The Trans-Voice App, used for transgender auto-evaluation, has a decision function based on a Multi-Layer Perceptron trained with a binary gender and arbitrary thresholds to obtain masculine, feminine, and androgynous voice categories. Its output was compared to speaker judgments on their own speech, with an accuracy of 88\%~\cite{williams22}.

Categorical systems make it difficult to monitor the speaker's progress during their transition. Our working assumption is to favor systems producing  continuous gender estimates fitted to human perception of gender. An LDA system based on 29 acoustic features was trained on cisgender voices annotated on a continuous scale~\cite{chen2020objective,CHEN202322}. While not addressing transgender voices, this work required excerpts of at least 7 seconds to obtain predictions correlated with perception and found that mean $F_0$, third and fourth formants, and vocal tract length (VTL) were the most correlated features with perceived gender.





\section{Cis- and transgender voices corpus}
\label{sec:corpus}
\subsection{Recording and analysis}



41 speakers were recorded reading the French version of The North Wind and the Sun.
They were 8 cisgender males (CM), 12 cisgender females (CF), and 21 transgender females (TF), with age varying between 20 and 69 years old (mean: 39). 
The TF speakers had transgender voice therapy supervised by one author of this study; none of them received surgical processing of the vocal apparatus.
All speakers signed an informed consent form detailing the aims of the research project to allow their voices to be used for research purposes only.
Recordings were made either at hospital Tenon AP-HP in a quiet room for transgender and some cisgender speakers or at the LISN laboratory for some cisgender voices. Recordings were made using a microphone at about 30 cm from the speaker's mouth, with a Nacon microphone at the hospital or a Zoom H4n recorder with its default microphones at the lab.
The readings had an average duration of 39 seconds, varying between 30 and 51 seconds.

$F_0$ was estimated following recommendations in \cite{vaysse2022}, combining \texttt{REAPER}'s voicing estimation~\cite{talkin2015reaper} with \texttt{FCN-F0}'s $F_0$ estimation~\cite{Ardaillon2019}. 
$F_0$ was expressed in semitones (ST) relative to 1 Hz.
Estimation of VTL was made using the first four formants (measured on the vocalic part of the readings), using \cite{Lammert_Narayanan_2015}'s equation and recommendations for formants estimation, using Praat's Burg algorithm \cite{Boersma_Weenink_2022} (i.e., estimating 6 formants with a 5.5kHz frequency threshold).



\subsection{Gender perception test}

A perception test was conducted using PsyToolkit \cite{Stoet2010,Stoet2016}, an online interface allowing the realization of in-browser experiments. A link to the online interface was sent to multiple French-speaking research mailing lists and  social media. 57 participants were enrolled in this perceptual evaluation. They were asked to provide their gender (35 female, 20 male, 2 other or confidential) and age range (18 in 18-35 years old, 25 in 36-50, 9 in 51-65, 4 over 65, 1 confidential).

Participants had to read the instructions and to accept participating in the study. Instructions described how this research aims at investigating why and how voices are perceived as produced by females or males and that the participants were supposed to evaluate how the voice they were about to listen to could have been produced by a female or a male, of a given age.  Participants were not told that the voices might contain transgender voices in order to avoid influencing their decisions. 
The 41 recordings were presented in a random order to each participant, who had to answer two questions intuitively and rapidly without having to listen to the whole speech sample. Participants had the possibility to answer "I don't know" (IDK). The questions were
Q1: What is the voice's gender? (answers: Female, Male, IDK); and 
Q2: How old is the speaker? (answers: 20-35, 36-50, 51-65, over 65, IDK).

Q1 answer buttons appeared at the beginning of the stimulus presentation. Participants had to answer Q1 to be able to answer Q2. They were not able to replay stimuli nor change their answers. Once the two answers were recorded, the next stimulus was presented after a short pause. The evaluation took about 6 minutes to complete, excluding the time required to read the instructions and provide demographic data.
The question related to speaker age (Q2) was aimed at distracting participants to avoid a focus on gender. 
For each question, the answers and the associated reaction times (RT) were recorded. The answers related to the speaker's age are not used in this study.

\subsection{Perceptual evaluation results}

Table \ref{tab:rawstats} presents the proportion of Q1 answers by speaker's categories. For cisgender speakers, a negligible amount of errors or IDK answers were observed (resp. 0.4 and 0.2\%) 
together with
shorter average RT: 3.4 and 3.7 seconds, versus 6.2 seconds for TF speakers. 
Participants tend to attribute a binary gender category to transgender voices, with a notable but modest increase (5\%) of IDK answers. 
Mean gender judgments are 0.539 for female listeners and 0.565 for male listeners.  Wilcoxon rank sum test shows no significant differences with probability $>= 95\%$ between these 2 groups (W = 268.5, p-value = 0.1548).

\begin{table}[th]
  \caption{Proportion of Q1 answer categories and Reaction Times (RT) by speaker category (CF, CM, TF).}
  \label{tab:rawstats}
  \centering
  \begin{tabular}{ l c c c }
    \toprule
     & \textbf{CF} & \textbf{CM} & \textbf{TF} \\
    \midrule
    Perceived as Female (\%) & 99.6 & 0 & 47.6 \\
    Perceived as Male (\%) & 0.4 & 99.8 & 47.4 \\
    IDK (\%) & 0 & 0.2 & 5.0 \\
    \midrule
    Average RT (s) & 3.4&  3.7 & 6.2 \\ 
    Standard deviation RT (s) & 4.1 &  4.3 & 5.8 \\
    \bottomrule
  \end{tabular}
\end{table}

\begin{figure}[th!]
  \centering
  \includegraphics[width=\linewidth]{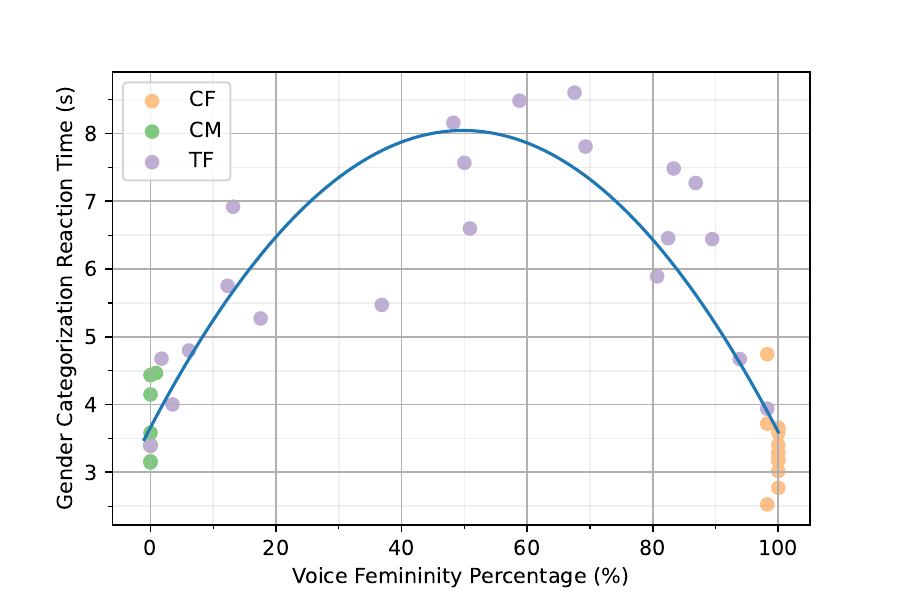}
  \caption{By speaker plot of the mean VFP by average RT, with a $2^{nd}$ order polynomial fit predicting RT from VFP}
  \label{fig:perceptionVSRT}
\end{figure}

From this perceptual evaluation, we defined a perceived "Voice Femininity Percentage" (VFP) index, derived from Q1 answers, and defined as the number of Female answers plus half of the IDK answers, divided by the total number of answers.
Figure \ref{fig:perceptionVSRT} shows the mean RT for each speaker's voice, according to their VFP.
While CM and CF speakers have VFP close to 0 and 100, TF VFPs are spread between 0 and 100.
A second-order polynomial gives a reasonable fit of the RT, as a function of voice VFP, with a maximal RT centered around 50\% VFP.


%
%




\section{Gender prediction models}



The complete voice gender evaluation system is based on three components that take as input a wav file sampled at 16~kHz. 
The first step is based on the \texttt{inaSpeechSegmenter} Voice Activity Detector used to discard non-speech segments~\cite{doukhan2018ina}.
Then, a 2D CNN gender classification model is applied using a sliding window. At each window step, it produces a binary gender prediction that is averaged over the complete recording.
Lastly, an Isotonic regression calibration procedure~\cite{scikit-learn,chakravarti1989isotonic} transforms this average score to the VFP obtained on the Trans- and cis-gender corpus using a non-linear increasing mapping. 

\subsection{Binary speaker gender training corpora}

Table \ref{tab:trainingcorpora} presents the 4 corpora 
used to train or evaluate binary gender classification models.
Voxceleb 2 (Vox2) contains celebrity voices obtained from Youtube in various recording conditions~\cite{Chung18b}. Despite its size, it mostly features English speakers, which may be sub-optimal for training gender detection systems targeting French voices.
INA's speaker dictionary (INA1) and diachronic speaker corpus (INA2) contain voices of celebrities obtained from French audiovisual archives~\cite{salmon2014effortless,uro2022semi}.
INA1 is based on speech broadcast on French TV news between 2007 and 2013.
INA2 subset contains TV and radio speech broadcast in 2015-2016, balanced across 4 age ranges (20-35, 36-50, 51-65, 65+) and 2 genders (female, male).
The French set of Common Voice corpus (CVFr) contains a large number of anonymous volunteer speakers reading short sentences in French and recorded using their own devices \cite{commonvoice:2020}. CVFr has interesting properties with respect to our final use case, where individuals may use variable-quality recording devices.

\begin{table}[th]
  \caption{Corpora used for training gender classification systems, described by number of unique female (\#F) and male (\#M) speakers, duration in hours (Dur), main language (Lang) and availability (Av)}
  \label{tab:trainingcorpora}
  \centering
  \begin{tabular}{ l c c c c c}
    \toprule
     \textbf{Corpus} & \textbf{\#F} & \textbf{\#M} & \textbf{Dur} &   \textbf{Lang} &  \textbf{Av}\\
    \midrule
    Vox2~\cite{Chung18b} & 2311 & 3682 & 2460  & English & Public\\
    INA1~\cite{salmon2014effortless}  & 494 & 1790 & 123  & French & Request\\
    INA2~\cite{uro2022semi} & 122 & 165 & 39 & French & Request\\
    CVFr~\cite{commonvoice:2020} & 758 & 3070 & 478 & French & Public\\
    \bottomrule
  \end{tabular}
\end{table}

\subsection{2D CNN speaker gender classification}

Two types of 2D CNN architectures were investigated for building the classification models. Both operate on Mel-scaled filterbank coefficients obtained from 25 ms windows with a step size of 10 ms.
2D CNN inputs are defined as \textit{patches} of dimensions T*N, with N the number of Mel bands extracted from each analysis window, and $T=150$ being the time dimension (the number of signal windows required to create an input patch, corresponding to 1515~ms speech excerpts). 





We defined Temporal Pooling CNN architectures (TpCnn) inspired by~\cite{doukhan2018open} using N=24 Mel-scaled filter bank inputs.
These architectures are based on NCONV convolutional blocks, a temporal pooling layer (\verb5<maxpool,T,15), NDENSE dense layers, and a sigmoid activation.
Convolutional blocks are composed of \textit{valid} K1*K2 kernels with NFILT filters followed by batch normalization and RELU activation.
Frequency (\verb5<maxpool,1,2>5), Time (\verb5<maxpool,2,1>5),  or Time-frequency (\verb5<maxpool,2,2>5) invariance pooling strategy were inserted between convolutions blocks.
Dense layers contain NN neurons and have Dropout rates of $0.2$. The parameter space of architectures was explored with NCONV varying from 2 to 5, NDENSE varying from 0 to 4, NFILT and NN in set $\{32, 64, 128, 256, 512\}$, K1 and K2 in set $\{3,5,7,9\}$ and varying pooling strategies.





We also defined several X-vector based architectures using VBX open-source extractor~\cite{vbx}.
X-vectors are 1-dimensional speaker embeddings obtained with DNN architectures, generally used for speaker-related tasks (recognition, verification, diarization)~\cite{snyder2018x}.
VBX extractor is based on a \texttt{Resnet101} architecture pre-trained on Vox2 corpus, using 64-dimensional Mel filterbank coefficients to obtain 256-dimensional X-vectors.
This deep model (347 layers) has a relatively large amount of parameters (45 M) since it was trained for complex tasks. 
We build on top of this extractor several Multi-Layer Perceptron (MLP),
with a number of layers varying between 1 and 4 and the number of neurons per layer in set $\{32, 64, 128, 256, 512\}$.

\subsection{Training Strategy}

We defined a DNN training strategy aimed at obtaining models with minimal gender, corpus, and speaker biases.
Male speakers were randomly excluded from corpora so to obtain balanced subsets containing the same amount of unique male and female speakers.
To mix training corpora, we discarded speakers from the largest in order to obtain subsets with the same amount of unique speakers per corpus.
Speech recordings were then grouped by unique speaker identifier and split into mutually exclusive training and development sets using ratios of 80 and 20\%, so a speaker from the train set is absent from the dev set.
For each epoch, a 1515 ms speech excerpt was randomly drawn (position, recording condition) for each speaker, resulting in a sample number equal to the number of unique speakers, balanced across genders and corpora.
Models were then trained using an early stopping procedure with patience set at 50 epochs, monitoring the estimate defined as the global loss plus the absolute value of the loss difference between male and female speakers obtained on the development set.
Each model was trained using 3 random initializations, and objective function convergence was obtained within a maximal amount of 160 epochs.
1500 TpCNN and 200 Xvector-based models were trained using NVIDIA 2080 Ti GPUs, requiring 850 hours of computation time (30 minutes/model).

\section{Results}

Evaluations were realized in a cross-corpus configuration.
Vox2, INA1, and CVFr corpora were used to train ML models in single and mixed corpus configurations (\texttt{French}=INA1+CvFr and \texttt{All}=Vox2+INA1+CvFr).
INA2 was used for testing models on the binary gender classification (BGC) task and to obtain estimates of accuracy per gender and age category.
The Trans- and Cisgender voice Corpus (TCC) was used for testing the Voice Femininity Percentage (VFP) prediction. 
Our proposals are compared to 4 baselines: \texttt{F0} and \texttt{VTL} corresponding to median $F_0$ or VTL, \texttt{F0VTL} is a linear SVM fit on median $F_0$ and VTL features, \texttt{ISS} is a gender classification model provided in the open-source project \texttt{inaSpeechSegmenter} and pre-trained on French data~\cite{doukhan2018open}. 
These baselines were used in pipelines, including VAD and isotonic calibration.

Table \ref{tab:resultssynth}  presents the best VFP prediction models.
VFP results are reported separately for cis- (CIS) and transgender (TF) speakers using the coefficient of determination ($R^2$) observed between model predictions and perceptual estimates.
Each model is associated with (i) a binary gender classification (BGC) performance metrics described as the harmonic mean of the accuracy obtained for male and female speakers ($Hacc$) and (ii) a Gender Bias ($GB$) defined as the difference between the accuracy for male minus for female speakers ($GB > 0$ if male accuracy is higher than female accuracy, else $<0$; $GB$ close to 0 is better).
While $F_0$- and VTL-based models allowed obtaining reasonable VFP results for cisgender speakers ($R^2=0.94$), their ability to predict transgender VFP is lower ($R^2=0.53$), 
illustrating the limitation of the $F_0$ and VTL features for predicting transgender voices' perceived femininity (or gender).
While showing better abilities to estimate TF VFP ($R^2=0.79$), the \texttt{ISS} baseline was associated with lower scores than our proposals and a large gender bias ($GB=+4.6$).

For all training set configurations, \texttt{TpCNN} obtained lower scores than X-vector architectures.
Reported \texttt{TpCNN} results are limited to their best training set configuration using all available training data (TF VFP $R^2=0.86$); their lowest results were obtained while trained on CvFr ($R^2=0.76$).
X-vector models obtained CIS VFP $R^2 > 0.99$ for all training configurations, corresponding to almost perfect VFP estimation for cisgender speakers. 
Best TF VFP was obtained with a model using four 512 neuron hidden layers on the top of the \texttt{X-vector} extractor, trained in a single corpus configuration using CvFr ($R^2=0.94$).
It was associated with the lowest reported BGC gender bias ($GB=0.1$) but also with the lowest BGC harmonic accuracy ($Hacc=94.2$).
Best BGC results were obtained with different settings: a single 512-neuron hidden layer MLP trained with all the available data ($Hacc=98.1$). 
This best BGC-performing model resulted in a lower but fair TF VFP prediction ($R^2=0.92$).
These two best-performing models (TF VFP and BGC) were associated with BGC harmonic accuracy decreasing with the speaker's age, as illustrated in table~\ref{tab:bestmodelsynth}.




\begin{table}[th]
  \caption{Best VFP prediction models obtained. $Hacc$ and $GB$ are the harmonic accuracy and the gender bias obtained on the binary gender classification task. VFP $R^2$ is reported for cis- (CIS) and transgender (TF) speakers.}
  \label{tab:resultssynth}
  \centering
  \begin{tabular}{l|c|cc|cc} 
    \toprule
     \textbf{Model}  &  \textbf{Training}   & \multicolumn{2}{c|}{\textbf{BGC}} & \multicolumn{2}{c}{\textbf{VFP} $\mathbf{R^2}$} \\
       &  \textbf{corpus}& Hacc & GB & CIS & TF \\ 
    \midrule
    \texttt{F0} &  TCC & & & 0.8923 & 0.4886  \\
    \texttt{VTL} &  TCC & & & 0.6961 & 0.0586 \\
    \texttt{F0VTL}  &  TCC & & & 0.9407 & 0.5303 \\
    \midrule
    \texttt{ISS} &  INA1 & 93.5 & +4.6 & 0.985 &  0.792 \\
    \midrule
    \texttt{TpCNN} 
    &  All & 94.8 & +2.1 & 0.9978 & 0.8586 \\
    \midrule
    \texttt{X-vector} 
    & Vox2 & 96.0 & +5.8 & 0.9997 & 0.9181\\ 
    & INA1 & 97.3 & +1.3 & 0.9995 & 0.9149\\ 
    & CvFr & 94.2 & \textbf{+0.1} & 0.9987 & \textbf{0.9420}\\ 
    & French   & 97.6 & +2.4 & 0.9998 & 0.9147\\ 
    & All  & \textbf{98.1} & +1.5 & 0.9997 & 0.9153\\ 
    \bottomrule
  \end{tabular}
\end{table}

\begin{table}[th]
  \caption{Binary Gender Harmonic Accuracy ($Hacc$) and Gender Bias ($GB$) described by gender and age categories of the best binary gender classification model (\texttt{X-vector All}) and the best VFP prediction model (\texttt{X-vector CvFr})}
  \label{tab:bestmodelsynth}
  \centering
  \begin{tabular}{ l c c c c c}
    \toprule
     \textbf{Model} & &\textbf{20-35} & \textbf{36-50} & \textbf{51-65} & \textbf{over 65}\\
    \midrule
    X-vector All & Hacc & 99.3 & 98.6 & 98.2 & 96.0\\
    & GB     & -1.0 & -0.6 & +3.1 & +4.3 \\
    \midrule 
    X-vector CvFr & Hacc & 96.2 & 95.6 & 94.3 & 90.3\\
    & GB     & -2.1 & -3.2 & +2.5 & +2.7\\
    \bottomrule
  \end{tabular}
\end{table}

\section{Conclusion}

We presented an original approach for estimating a continuous ratio of perceived gender from voice, defined as a Voice Femininity Percentage, and fitted to the perceptual results of a group of French speakers. 
This approach differs from~\cite{williams22}, as we have chosen to base our estimates on external listener judgments rather than on speakers' own judgments, as the former better fits our aim: reflecting the gender perceived by the interlocutors.
Unlike ~\cite{chen2020objective,CHEN202322}, we asked perceptual test participants to provide binary gender judgments because gender is mostly perceived as a binary characteristic in the French society (as shown by the barely used IDK option) -- but we considered the \textit{proportion} of female answers, that allowed us working on a continuous dimension. 
While we considered this perception task more natural than asking for continuous gender judgments, it required a significant group of participants; resulting in costly perceptual gender estimations that we considered necessary with respect to our final use case -- having a model that reflects how a voice would be perceived in a social interaction setting.

We implemented several machine learning models in charge of reproducing these perceptual judgments and obtained convincing results for cisgender ($R^2>0.99$) and transgender voices ($R^2=0.94$), which were shown to be much more accurate than predictions based on $F_0$ and/or VTL estimates only.
Best results were obtained using VBX X-vector features (pre-trained with English data)~\cite{vbx} with an MLP trained in a single corpus configuration using CvFr~\cite{commonvoice:2020}. 
This result suggests that the best performances were linked to speaking style similarity between CvFr and the evaluation material (non-professional read speech) rather than to the training data language (no major differences between INA1 and Vox2) or the sheer size of the dataset (CvFr is smaller than Vox2). 
Additional work would be necessary to estimate the potential impact of training data language if style is controlled for, using X-vector extractors trained on French data.
This result also suggests that the trained models that obtained fair but not the best results on our evaluation task may be better suited to the analysis of spontaneous speech, which was not represented in our evaluation material. 
Additional work is necessary to constitute a spontaneous speech corpus using similar gender perceptual evaluation protocols.
Other factors, and typically the speaker's age, had a major effect on all models evaluation metrics, and typically gender bias: these results may reflect literature describing the evolution of voice with age during adulthood \cite{Sataloff_Kost_Linville_2017,Yamauchi_Yokonishi_Imagawa_Sakakibara_Nito_Tayama_Yamasoba_2015}, with decreased $F_0$ in female vs. an increase for male.
This illustrates the importance of models fitted to the voice of speakers with varied characteristics.

Results described in this study are currently limited to read speech in French.
Ongoing work consists in building Human-Machine Interfaces to investigate if these theoretical results match end-users expectations and allow to provide constructive voice-passing feedback to be used in addition or instead of $F_0$ estimates.
%
Best performing BGC models presented in this study have been integrated to \texttt{inaSpeechSegmenter} open-source project~\cite{anonymous}. 
Discussions with French regulatory authorities are necessary to define how fitted calibration modules (BGC to VFP mapping) could be disseminated while preventing non-ethical uses 
related to the characterization of non-prototypical voices. 


\section{Acknowledgements}

\ifinterspeechfinal
     This work has been partially funded by the French National Research Agency (project Gender Equality Monitor - ANR-19-CE38-0012).
\else
     For anonymity issues, authors will describe upon acceptance how this project has been funded together with instructions for obtaining the software.
\fi


\bibliographystyle{IEEEtran}
\bibliography{mybib}

\begin{thebibliography}{10}
\providecommand{\url}[1]{#1}
\csname url@samestyle\endcsname
\providecommand{\newblock}{\relax}
\providecommand{\bibinfo}[2]{#2}
\providecommand{\BIBentrySTDinterwordspacing}{\spaceskip=0pt\relax}
\providecommand{\BIBentryALTinterwordstretchfactor}{4}
\providecommand{\BIBentryALTinterwordspacing}{\spaceskip=\fontdimen2\font plus
\BIBentryALTinterwordstretchfactor\fontdimen3\font minus
  \fontdimen4\font\relax}
\providecommand{\BIBforeignlanguage}[2]{{%
\expandafter\ifx\csname l@#1\endcsname\relax
\typeout{** WARNING: IEEEtran.bst: No hyphenation pattern has been}%
\typeout{** loaded for the language `#1'. Using the pattern for}%
\typeout{** the default language instead.}%
\else
\language=\csname l@#1\endcsname
\fi
#2}}
\providecommand{\BIBdecl}{\relax}
\BIBdecl

\bibitem{gray2019transgender}
M.~L. Gray and M.~S. Courey, ``Transgender voice and communication,''
  \emph{Otolaryngologic Clinics of North America}, vol.~52, no.~4, pp.
  713--722, 2019.

\bibitem{Fugain_2019}
C.~Fugain, \emph{\BIBforeignlanguage{fre}{La puberté, la mue et la
  transidentité}}.\hskip 1em plus 0.5em minus 0.4em\relax IsBergues: Ortho
  Edition, 2019.

\bibitem{schmidt2018voice}
J.~G. Schmidt, B.~N. G.~d. Goulart, M.~E. K.~Y. Dorfman, G.~Kuhl, and L.~M.
  Paniagua, ``Voice challenge in transgender women: trans women self-perception
  of voice handicap as compared to gender perception of na{\"\i}ve listeners,''
  \emph{Revista CEFAC}, vol.~20, pp. 79--86, 2018.

\bibitem{murry1980multidimensional}
T.~Murry and S.~Singh, ``Multidimensional analysis of male and female voices,''
  \emph{The journal of the Acoustical society of America}, vol.~68, no.~5, pp.
  1294--1300, 1980.

\bibitem{hillenbrand2009role}
J.~M. Hillenbrand and M.~J. Clark, ``The role of f0 and formant frequencies in
  distinguishing the voices of men and women,'' \emph{Attention, Perception, \&
  Psychophysics}, vol.~71, pp. 1150--1166, 2009.

\bibitem{VanBorsel_VanEynde_DeCuypere_Bonte_2008}
J.~Van~Borsel, E.~Van~Eynde, G.~De~Cuypere, and K.~Bonte,
  ``\BIBforeignlanguage{en}{Feminine after cricothyroid approximation?}''
  \emph{\BIBforeignlanguage{en}{Journal of Voice}}, vol.~22, no.~3, p.
  379–384, May 2008.

\bibitem{Mastronikolis_Remacle_Biagini_Kiagiadaki_Lawson_2013}
N.~S. Mastronikolis, M.~Remacle, M.~Biagini, D.~Kiagiadaki, and G.~Lawson,
  ``\BIBforeignlanguage{en}{Wendler glottoplasty: An effective pitch raising
  surgery in male-to-female transsexuals},''
  \emph{\BIBforeignlanguage{en}{Journal of Voice}}, vol.~27, no.~4, p.
  516–522, Jul 2013.

\bibitem{evaf}
\BIBentryALTinterwordspacing
{VoxPop, LLC}. {EvaF} : Voice training tools \& lessons. [Online]. Available:
  \url{https://www.evaf.app}
\BIBentrySTDinterwordspacing

\bibitem{voiceup}
\BIBentryALTinterwordspacing
{Speechtools Ltd}. {Christella VoiceUp} : Trans woman voice training. [Online].
  Available:
  \url{http://www.christellaantoni.co.uk/transgender-voice/voiceupapp}
\BIBentrySTDinterwordspacing

\bibitem{vanBezooijen_1995}
R.~van Bezooijen, ``\BIBforeignlanguage{en}{Sociocultural aspects of pitch
  differences between japanese and dutch women},''
  \emph{\BIBforeignlanguage{en}{Language and Speech}}, vol.~38, no.~3, p.
  253–265, Jul 1995.

\bibitem{childers88}
D.~Childers, K.~Wu, K.~Bae, and D.~Hicks, ``Automatic recognition of gender by
  voice,'' in \emph{ICASSP-88., International Conference on Acoustics, Speech,
  and Signal Processing}, 1988, pp. 603--606.

\bibitem{lamel1995phone}
L.~Lamel and J.-L. Gauvain, ``A phone-based approach to non-linguistic speech
  feature identification,'' \emph{Computer Speech \& Language}, vol.~9, no.~1,
  1995.

\bibitem{543213}
E.~Parris and M.~Carey, ``Language independent gender identification,'' in
  \emph{1996 IEEE International Conference on Acoustics, Speech, and Signal
  Processing Conference Proceedings}, vol.~2, 1996, pp. 685--688 vol. 2.

\bibitem{lebourdais2022overlaps}
M.~Lebourdais, M.~Tahon, A.~Laurent, S.~Meignier, and A.~Larcher, ``Overlaps
  and gender analysis in the context of broadcast media,'' in \emph{Proceedings
  of the Thirteenth Language Resources and Evaluation Conference}, 2022, pp.
  3264--3270.

\bibitem{doukhan2018open}
D.~Doukhan, J.~Carrive, F.~Vallet, A.~Larcher, and S.~Meignier, ``An
  open-source speaker gender detection framework for monitoring gender
  equality,'' in \emph{2018 IEEE international conference on acoustics, speech
  and signal processing (ICASSP)}.\hskip 1em plus 0.5em minus 0.4em\relax IEEE,
  2018, pp. 5214--5218.

\bibitem{bensoussan2021deep}
Y.~Bensoussan, J.~Pinto, M.~Crowson, P.~R. Walden, F.~Rudzicz, and
  M.~Johns~III, ``Deep learning for voice gender identification:
  proof-of-concept for gender-affirming voice care,'' \emph{The Laryngoscope},
  vol. 131, no.~5, pp. E1611--E1615, 2021.

\bibitem{yasmin2022rough}
G.~Yasmin, A.~K. Das, J.~Nayak, S.~Vimal, and S.~Dutta, ``A rough set theory
  and deep learning-based predictive system for gender recognition using audio
  speech,'' \emph{Soft Computing}, pp. 1--24, 2022.

\bibitem{williams22}
J.~Williams and P.~Paudel, ``Application of deep feedforward neural network in
  transgender vocal analysis,'' St. Olaf College, Northfield, Minnesota,
  U.S.A., Tech. Rep., 2022.

\bibitem{chen2020objective}
F.~Chen, R.~Togneri, M.~Maybery, and D.~Tan, ``An objective voice gender
  scoring system and identification of the salient acoustic measures.'' in
  \emph{INTERSPEECH}, 2020, pp. 1848--1852.

\bibitem{CHEN202322}
F.~Chen, R.~Togneri, M.~Maybery, and D.~W. Tan, ``Acoustic characterization and
  machine prediction of perceived masculinity and femininity in adults,''
  \emph{Speech Communication}, vol. 147, pp. 22--40, 2023.

\bibitem{vaysse2022}
R.~Vaysse, C.~Ast\'esano, and J.~Farinas, ``Performance analysis of various
  fundamental frequency estimation algorithms in the context of pathological
  speech,'' \emph{The Journal of the Acoustical Society of America}, vol. 152,
  no.~5, pp. 3091--3101, 2022.

\bibitem{talkin2015reaper}
\BIBentryALTinterwordspacing
D.~Talkin, ``{REAPER}: Robust epoch and pitch estimator,'' 2015. [Online].
  Available: \url{https://github.com/google/REAPER}
\BIBentrySTDinterwordspacing

\bibitem{Ardaillon2019}
L.~Ardaillon and A.~Roebel, ``{Fully-Convolutional Network for Pitch Estimation
  of Speech Signals},'' in \emph{Proc. Interspeech 2019}, 2019, pp. 2005--2009.

\bibitem{Lammert_Narayanan_2015}
A.~C. Lammert and S.~S. Narayanan, ``\BIBforeignlanguage{en}{On short-time
  estimation of vocal tract length from formant frequencies},''
  \emph{\BIBforeignlanguage{en}{PLOS ONE}}, vol.~10, no.~7, p. e0132193, Jul
  2015.

\bibitem{Boersma_Weenink_2022}
\BIBentryALTinterwordspacing
P.~Boersma and D.~Weenink, ``Praat: doing phonetics by computer [computer
  program]. version 6.2.08,'' Feb 2022. [Online]. Available:
  \url{http://www.praat.org/}
\BIBentrySTDinterwordspacing

\bibitem{Stoet2010}
G.~Stoet, ``{PsyToolkit}: A software package for programming psychological
  experiments using linux,'' \emph{Behavior Research Methods}, vol.~42, no.~4,
  pp. 1096--1104, Nov. 2010.

\bibitem{Stoet2016}
------, ``{PsyToolkit},'' \emph{Teaching of Psychology}, vol.~44, no.~1, pp.
  24--31, Nov. 2016.

\bibitem{doukhan2018ina}
D.~Doukhan, E.~Lechapt, M.~Evrard, and J.~Carrive, ``Ina’s mirex 2018 music
  and speech detection system,'' \emph{Music Information Retrieval Evaluation
  eXchange (MIREX 2018)}, 2018.

\bibitem{scikit-learn}
F.~Pedregosa, G.~Varoquaux, A.~Gramfort, V.~Michel, B.~Thirion, O.~Grisel,
  M.~Blondel, P.~Prettenhofer, R.~Weiss, V.~Dubourg, J.~Vanderplas, A.~Passos,
  D.~Cournapeau, M.~Brucher, M.~Perrot, and E.~Duchesnay, ``Scikit-learn:
  Machine learning in {P}ython,'' \emph{Journal of Machine Learning Research},
  vol.~12, pp. 2825--2830, 2011.

\bibitem{chakravarti1989isotonic}
N.~Chakravarti, ``Isotonic median regression: a linear programming approach,''
  \emph{Mathematics of operations research}, vol.~14, no.~2, pp. 303--308,
  1989.

\bibitem{Chung18b}
J.~S. Chung, A.~Nagrani, and A.~Zisserman, ``Voxceleb2: Deep speaker
  recognition,'' in \emph{INTERSPEECH}, 2018.

\bibitem{salmon2014effortless}
F.~Salmon and F.~Vallet, ``An effortless way to create large-scale datasets for
  famous speakers.'' in \emph{LREC}, 2014, pp. 348--352.

\bibitem{uro2022semi}
R.~Uro, D.~Doukhan, A.~Rilliard, L.~Larcher, A.-C. Adgharouamane, M.~Tahon, and
  A.~Laurent, ``A semi-automatic approach to create large gender-and
  age-balanced speaker corpora: Usefulness of speaker diarization \&
  identification,'' in \emph{13th Language Resources and Evaluation
  Conference}, 2022, pp. 3271--3280.

\bibitem{commonvoice:2020}
R.~Ardila, M.~Branson, K.~Davis, M.~Henretty, M.~Kohler, J.~Meyer, R.~Morais,
  L.~Saunders, F.~M. Tyers, and G.~Weber, ``Common voice: A
  massively-multilingual speech corpus,'' in \emph{Proceedings of the 12th
  Conference on Language Resources and Evaluation (LREC 2020)}, 2020, pp.
  4211--4215.

\bibitem{vbx}
\BIBentryALTinterwordspacing
F.~Landini, J.~Profant, M.~Diez, and L.~Burget, ``Bayesian hmm clustering of
  x-vector sequences (vbx) in speaker diarization: theory, implementation and
  analysis on standard tasks,'' 2020. [Online]. Available:
  \url{https://arxiv.org/abs/2012.14952}
\BIBentrySTDinterwordspacing

\bibitem{snyder2018x}
D.~Snyder, D.~Garcia-Romero, G.~Sell, D.~Povey, and S.~Khudanpur, ``X-vectors:
  Robust dnn embeddings for speaker recognition,'' in \emph{2018 IEEE
  international conference on acoustics, speech and signal processing
  (ICASSP)}.\hskip 1em plus 0.5em minus 0.4em\relax IEEE, 2018, pp. 5329--5333.

\bibitem{Sataloff_Kost_Linville_2017}
R.~T. Sataloff, K.~M. Kost, and S.~E. Linville, \emph{Chapter 13. The Effects
  of Age on the Voice}, second edition~ed.\hskip 1em plus 0.5em minus
  0.4em\relax San Diego, CA: Plural Publishing, Inc, 2017, p. 221–240.

\bibitem{Yamauchi_Yokonishi_Imagawa_Sakakibara_Nito_Tayama_Yamasoba_2015}
A.~Yamauchi, H.~Yokonishi, H.~Imagawa, K.-I. Sakakibara, T.~Nito, N.~Tayama,
  and T.~Yamasoba, ``\BIBforeignlanguage{en}{Quantitative analysis of digital
  videokymography: A preliminary study on age- and gender-related difference of
  vocal fold vibration in normal speakers},''
  \emph{\BIBforeignlanguage{en}{Journal of Voice}}, vol.~29, no.~1, p.
  109–119, Jan 2015.

\bibitem{anonymous}
\BIBentryALTinterwordspacing
D.~Doukhan, ``{inaSpeechSegmenter} : a cnn-based audio segmentation toolkit,''
  2018. [Online]. Available:
  \url{https://github.com/ina-foss/inaSpeechSegmenter}
\BIBentrySTDinterwordspacing

\end{thebibliography}

\end{document}